\documentstyle[12pt,aasms4]{article}

\newcommand \hii  {H\,{\sc ii}}

\newcommand \hi  {H\,{\sc i}}
\newcommand \ha   {H$\alpha$}

\newcommand \sii  {[S\,{\sc ii}]}

\begin{document}

\title{{\em HST} WFPC2 Imaging of Shocks in Superbubbles}

\author{C.-H. Rosie Chen, You-Hua Chu, Robert A. Gruendl, and Sean D. Points}
\affil{Astronomy Department, University of Illinois, 1002 W. Green Street,
Urbana, IL 61801 \\
Electronic-mail: c-chen@astro.uiuc.edu, chu@astro.uiuc.edu,
gruendl@astro.uiuc.edu, points@astro.uiuc.edu}

\begin{abstract}

Bright X-ray emission has been detected in superbubbles in the Large
Magellanic Cloud (LMC), and it is suggested that supernova remnants 
(SNRs) near the inner shell walls are responsible for this X-ray 
emission.  To identify SNR shocks in superbubble interiors, we have 
obtained {\em HST} WFPC2 emission-line images of the X-ray-bright 
superbubbles DEM\,L\,152 and DEM\,L\,192 and the X-ray-dim superbubble 
DEM\,L\,106.  We use these images to examine the shell morphology and 
\sii/\ha\ ratio variations in detail.

Of these three superbubbles, DEM\,L\,152 has the highest X-ray surface 
brightness, the most filamentary nebular morphology, the 
largest expansion velocity ($\sim$40 km~s$^{-1}$), and the highest 
\sii/\ha\ ratio (0.4--0.6).  Its \sii/\ha\ ratio increases outwards 
and peaks in sharp filaments along the periphery.  
DEM\,L\,192 has a moderate X-ray surface brightness, a complex but not 
filamentary morphology, a moderate expansion velocity (35 km~s$^{-1}$), 
and a low \sii/\ha\ ratio ($\sim$0.15).  
DEM\,L\,106 is not detected in X-rays.  Its shell structure is amorphous 
and has embedded dusty features; its expansion velocity is 
$<$10 km~s$^{-1}$.

None of the three superbubbles show morphological features in the
shell interior that can be identified as directly associated with
SNR shocks, indicating that the SNR shocks have not encountered 
very dense material.  We find that the \sii/\ha\ ratios of 
X-ray-bright superbubbles are strongly dependent on the UV radiation 
field of the encompassed OB associations.  Therefore, a tight 
correlation between \sii/\ha\ ratio and X-ray surface brightness in 
superbubbles should not exist.  We also find that the filamentary 
morphologies of superbubbles are associated with large expansion 
velocities and bright X-ray emission.

\end{abstract}


\clearpage


\section{Introduction: SNRs in Superbubbles}

Bright X-ray emission has been detected in superbubbles in the Large Magellanic
Cloud (LMC).  As their X-ray luminosities are much higher than those expected 
in superbubble models (e.g., \markcite{We77}Weaver et al.\ 1977), it is suggested
that supernova remnants (SNRs) near the inner shell walls are responsible 
for the X-ray emission \markcite{CM90}(Chu \& Mac Low 1990; \markcite{WH91}Wang
\& Helfand 1991). 

SNRs in superbubbles cannot be confirmed easily by the conventional diagnostics
of nonthermal radio emission and a high \sii/\ha\ line ratio. The radio and 
optical emission of SNRs in superbubbles is weak because the SNRs interact 
with a low-density medium.  Their weak nonthermal radio emission is further 
drowned out by thermal emission from bright background \hii\ regions.  The 
\sii\ line strength is weakened because sulfur may be photoionized to higher 
ionization stages by the UV radiation from the OB associations in the 
superbubble.  Therefore, the conventional methods for identifying SNRs are 
ineffective for SNRs in superbubbles. 

Alternative methods have been used to search for evidence of SNRs in 
X-ray-bright superbubbles.  One of these attempts has used high-dispersion 
spectroscopic observations of the \ha\ emission line to search for 
high-velocity ($\Delta V \geq 100$ km~s$^{-1}$) shocked gas in X-ray-bright 
superbubbles.  The emission measures of the shocked gas, as derived from 
the X-ray surface brightness, are high, but high-velocity gas is not 
detected in the 4m echelle/CCD observations \markcite{C97}(Chu 1997).  
This negative result is consistent with the suggestion that the SNR shocks 
are interacting with the hot, low-density interior of the superbubble.  
The small recombination coefficient and large thermal width associated 
with the high temperatures in the post-shock gas prohibit the detection 
of high-velocity gas in the \ha\ emission line profiles.

One other attempt to search for SNRs in superbubble interiors used the 
UV interstellar absorption line properties.  If the SNR shocks are 
interacting mainly with the hot interior of a superbubble and have not 
yet reached the cold, dense shell wall, the high-ionization species will 
exhibit different velocity profiles than the low-ionization species.  
The high-ionization species are expected to possess an additional shocked, 
high-velocity component.  \markcite{Ch94}Chu et al.\ (1994) examined all 
available archival high-dispersion {\em International Ultraviolet Explorer}
({\em IUE}) spectra of targets in the LMC, and found promising diagnostics 
of SNR shocks only in the superbubble N51D \markcite{DN82}(de Boer \& Nash 
1982) and the giant \hii\ region 30 Doradus.

Thus, we need to find other diagnostics to confirm the existence of SNRs 
in superbubbles.  As the interstellar medium is most likely clumpy, there 
might be dense cloudlets left in the superbubble interior.  Small, 
shocked cloudlets ($\le 1\arcsec $) cannot be resolved by ground-based 
telescopes, and hence would be difficult to identify.  The Wide Field 
and Planetary Camera 2 (WFPC2) on board the {\em Hubble Space Telescope} 
({\em HST}), being able to resolve features as small as 0\farcs2, or
0.05 pc for a distance of 50 kpc to the LMC \markcite{Fe91}(Feast 1991),
provides an opportunity to search for evidence of SNR shocks in 
superbubble interiors.  Indeed, small shocked cloudlets have been
successfully detected by {\em HST} WFPC2 images in the SNR N63A
\markcite{Ch99}(Chu et al.\ 1999).

We have selected three superbubbles of different X-ray surface 
brightnesses for {\em HST} WFPC2 imaging: DEM\,L\,106, DEM\,L\,152,
and DEM\,L\,192 in the LMC \markcite{DEM}(``DEM'' -- Davies, Elliott, 
\& Meaburn 1976).  DEM\,L\,152, part of N44 (``N'' -- Henize 1956), 
is the brightest X-ray superbubble in the LMC, with an X-ray luminosity 
L$_{\rm X} \sim$ a few $\times 10^{36}$ ergs~s$^{-1}$ in the 
0.5--2.0 keV band \markcite{Ch93}(Chu et al.\ 1993).  
DEM\,L\,192, also known as N51D, is of moderate X-ray surface brightness, 
with L$_{\rm X} \sim$ a few $\times 10^{35}$ ergs~s$^{-1}$ in the
0.15--4.5 keV band \markcite{CM90}(Chu \& Mac Low 1990).
DEM\,L\,106, or N30C, has the lowest X-ray surface brightness and has not 
been detected by {\it Einstein} or {\it ROSAT}; the 3$\sigma$ upper 
limit on its X-ray luminosity is L$_{\rm X} \le 10^{35}$ ergs~s$^{-1}$
in the 0.1--2.4 keV band \markcite{Ch95}(Chu et al.\ 1995).  These 
three superbubbles allow us to seek a possible relationship between the 
nebular morphology of the 10$^4$ K gas and the X-ray surface brightness 
of the 10$^6$ K gas, in order to determine how interior SNRs affect the
physical properties of a superbubble.

This paper reports the WFPC2 images of these three superbubbles.
In \S2, we describe the observations and data reduction.  In \S3, 
we describe the morphologies of the superbubbles and discuss the 
apparent relationship between their \ha\ morphologies and X-ray surface 
brightnesses.  We further compare these three superbubbles to other 
LMC fields of which archival {\it HST} WFPC2 \ha\ images are available,
and discuss whether the apparent relationship between \ha\ morphology 
and X-ray surface brightness can be generalized.  A summary of our 
conclusions is given in \S4.


\section{Observations and Data Reduction}

{\em HST} WFPC2 CCD images of DEM\,L\,106, DEM\,L\,152, and DEM\,L\,192 
were taken between 
1998 November and 1999 January for the Cycle 6 program 6698.  The journal 
of observations is given in Table 1.  Two filters were used, the \ha\ 
filter (F656N) and the \sii\ filter (F673N), so that we could search for 
small shocked features and further use the \sii/\ha\ ratio to probe their 
nature.  No \sii\ images were taken for DEM\,L\,106 because 
its surface brightness 
was low and its \ha\ observations required an entire orbit. Figure 1 shows 
these WFPC2 \ha\ images and their corresponding ground-based \ha\ images.

The images were processed using the standard $HST$ data pipeline and 
combined to remove cosmic rays.  To extract fluxes from the \ha\ and \sii\
images, we have followed the procedures for narrow-band 
photometry\footnote{See instructions at
http://www.stsci.edu/instruments/wfpc2/Wfpc2\_faq/wfpc2\_nrw\_phot\_faq.html.}.
We first divide a combined image by its total exposure time to obtain a 
count-rate map, and multiple it by the parameter PHOTFLAM in the image 
header to convert from count rates to flux densities.  We then use the 
task SYNPHOT in the STSDAS software package to determine filter widths, 
28.3 \AA\ for \ha\ and 63.3 \AA\ for \sii.  Finally, we multiply the flux 
densities by the corresponding filter widths to obtain fluxes. 

The flux-calibrated \sii\ images are divided by the flux-calibrated 
\ha\ images to derive \sii/\ha\ ratio maps.  To suppress the large 
\sii/\ha\ ratio fluctuations in faint (and hence noisy) regions, 
we have clipped pixels with fluxes less than 3$\sigma$ above the sky 
background for both the \ha\ and \sii\ images of DEM\,L\,152 and 
DEM\,L\,192.  The \sii/\ha\ ratio maps of DEM\,L\,152 and DEM\,L\,192 
are presented in Figure 2.
 
We have compared these \sii/\ha\ ratios of DEM\,L\,152 and DEM\,L\,192
to those derived from long-slit, low-dispersion CCD spectra taken with 
the 1.5-m telescope and the 2D-Frutti spectra taken with the 1-m telescope 
at Cerro Tololo Inter-American Observatory (\markcite{Y2K}Kennicutt 2000).
The two sets of observations were made at the same slit positions and
with a 5$''$-wide slit.  Spectra were extracted over slit lengths of 
$\sim20''$ in order to obtain adequate S/N ratios.  Two spectra were
extrcted for DEM\,L\,152 and four for DEM\,L\,192.  The slit positions 
were recovered by matching the surface brightness profiles of the \ha\ 
line along the slits to those of the WFPC2 \ha\ images.  Despite the 
coarse spatial resolution of the spectra, we find that the slit positions 
relative to the WFPC2 images can be determined to better than 1$''$, as 
the WFPC2 surface brightness profiles change significantly with a shift 
larger than 1$''$.  The \sii/\ha\ ratios derived from the CCD spectra 
agree with those derived from the 2D-Frutti spectra with a $\sim$20\% 
scatter.  The \sii/\ha\ ratios derived from the WFPC2 images are 
systematically higher than the spectroscopic values, 10\% higher for 
DEM\,L\,152 and 40\% higher for DEM\,L\,192.  This discrepancy might be 
caused by diffuse continuum emission in the WFPC2 images.  
This discrepancy does not affect our evaluation of the \sii/\ha\ 
variations within each superbubble.


\section{Results and Discussions}

The {\it HST} WFPC2 images allow us to examine the nebular 
morphologies and \sii/\ha\ ratios at high spatial resolution.  
Note that the \sii/\ha\ ratio, while often used as a shock 
indicator, is not uniquely related to the shock
velocity.  The \sii/\ha\ ratio is high at both low shock velocities 
($<$50 km~s$^{-1}$) and high shock velocities ($>$100 km~s$^{-1}$) 
(\markcite{SM79}Shull \& McKee 1979).  Furthermore, the \sii/\ha\ 
ratio is affected by the local UV radiation field.  If early-type
O stars exist in the vicinity, their UV radiation will ionize 
sulfur to S$^{+2}$ and S$^{+3}$ and weaken the \sii/\ha\ ratio
behind the shocks.

Below we describe the morphology and \sii/\ha\ ratios of each 
superbubble, and interpret the observations using our best knowledge
of the expansion velocity and stellar content of the superbubble.  
The superbubbles are presented in the order of decreasing X-ray 
luminosity and surface brightness.
Later, we discuss the relationship between the nebular 
morphology and the X-ray surface brightness, and suggest an 
interpretation.

\subsection{DEM\,L\,152 (N44)}

Among the three superbubbles studied, DEM\,L\,152 has the highest 
X-ray luminosity, L$_{\rm X}$ = 1--35 $\times 10^{36}$ 
ergs~s$^{-1}$ in the 0.5--2.0 keV band \markcite{Ch93}(Chu et al.\ 
1993; \markcite{Ma96}Magnier et al.\ 
1996)\footnote{The uncertainty in the X-ray luminosity is caused by
the uncertainty in the spectral fits and a large absorption correction 
at this soft energy band.  Chu et al.\ (1993) used only $ROSAT$ PSPC
observations for spectral fits, and found plasma temperatures of 
$1.5-3\times10^6$ K and absorption column densities of 10$^{22}$ cm$^{-2}$.
They derived an X-ray luminosity of $3-35\times10^{36}$ erg~s$^{-1}$ in the
0.5--2.0 keV band.  Magnier et al.\ (1996) used both $ROSAT$ and ASCA 
observations for spectral fits and found a plasma temperature of 
$6\times10^6$ K and an absorption column density of 10$^{21}$ cm$^{-2}$.
Using the spectral fits from Magnier et al., we calculated an X-ray
luminosity of $\sim1\times10^{36}$ erg~s$^{-1}$ in the 0.5--2.0 keV band.},
and the largest expansion 
velocity, $\sim$40 km~s$^{-1}$ (\markcite{ML91}Meaburn \& Laspias 1991; 
\markcite{C97}Chu 1997).  The X-ray 
emission from DEM\,L\,152 extends from its interior to beyond the southern 
rim, suggesting a breakout in the superbubble (Chu et al.\ 1993).  
This breakout has been confirmed by velocity structures of the \ha\ 
line and variations in the plasma temperature (\markcite{Ma96}Magnier 
et al.\ 1996).  

As shown in Figure 1a, DEM\,L\,152 has a very filamentary optical morphology. 
The WFPC2 field of view is centered near the breakout region.  The
WFPC2 \ha\ image indeed shows an apparent gap in the southern rim
at $5^{\rm h}22^{\rm m}33^{\rm s}$, $-$67$^{\circ}$58$'$12$''$ (J2000).  
Most of the filaments are straight and narrow ($\sim1''$ wide, 
$\sim$10--20$''$ long), and are parallel to the direction of the 
outflow.  The narrowest filaments have FWHM of $\sim$0\farcs5, or 0.12 pc. 

The \sii/\ha\ ratio map of DEM\,L\,152 (Figure 2a) shows that the \sii/\ha\
ratios of DEM\,L\,152 are generally in the range of 0.4--0.6, which is 
higher than those of normal \hii\ regions, but still lower than the
values typically seen in SNRs, $\ge0.7$.  Systematic variations of
the \sii/\ha\ ratios are seen.  First, the superbubble shell shows
lower \sii/\ha\ ratios, $\sim$0.4 (light blue), along the inner edge,
and higher \sii/\ha\ ratios, up to $\sim$0.6 (yellow), along the outer
edge.  Second, the long filaments along the outflow direction have 
higher \sii/\ha\ ratios, $\sim$0.5 (red to yellow).

The \sii/\ha\ ratios of DEM\,L\,152 can be explained by the radiation field
provided by the stars in the superbubble.  Two OB associations have been 
identified in the vicinity of DEM\,L\,152: LH47 in the superbubble's interior 
and western rim, and LH48 in the compact \hii\ region on the superbubble's
northern rim (\markcite{LH70}Lucke \& Hodge 1970; Chu \& Mac Low 1990).  
The earliest-type stars, earlier than O6, are located in either 
LH48 or the western edge of LH47, both are inside dense \hii\ regions.  
The stars that are projected in the central cavity of DEM\,L\,152 are all of 
later types, such as late-O or early-B (\markcite{OM95}Oey \& Massey 1995).
These relatively late early-type stars do not have a strong UV radiation 
field.  The weakness of the UV radiation field is testified by the presence 
of a neutral \hi\ shell exterior to the ionized gas shell of DEM\,L\,152, 
indicating that the ionization front is trapped within the superbubble 
shell \markcite{kim98a}(Kim et al.\ 1998a).  At the ionization front, 
sulfur is mostly singly ionized; therefore, the observed \sii/\ha\ ratios 
are high and peak at the outer 
edge of the ionized gas shell.

\subsection{DEM\,L\,192 (N51D)}

N51D has a moderate X-ray luminosity, L$_{\rm X} \sim$ $3\times 
10^{35}$ ergs~s$^{-1}$ in the 0.15--4.5 keV band; 
its X-ray emission is confined within the optical 
shell, with the brightest X-ray emission adjacent to the bright
eastern rim (\markcite{CM90}Chu \& Mac Low 1990).  The expansion
velocity of DEM\,L\,192 is moderate, $\sim$ 30--35 km~s$^{-1}$ 
(\markcite{La80}Lasker 1980; \markcite{MT80}Meaburn \& Terrett 1980).

Our WFPC2 images of DEM\,L\,192 are centered near the eastern rim, where 
the X-ray emission peaks.  Figure 1b shows that it has a complex but 
not filamentary optical morphology.  The only sharp narrow feature
occurs at the edge of the WF4 CCD, the western edge of the field of 
view.  This sharp feature appears to be the ionized surface of a
dense, irregular cloud.

As shown in Figure 2b, the \sii/\ha\ ratios in DEM\,L\,192 are in the 
range of 0.1--0.25 (green -- blue), much lower than those in DEM\,L\,152.
If the spectroscopically determined \sii/\ha\ ratios are used, the
\sii/\ha\ ratios of DEM\,L\,192 are 40\% lower, and the difference 
between DEM\,L\,192 and DEM\,L\,152 is even larger.
The sharp feature at the western edge has the highest \sii/\ha\ ratio,
up to $\sim$0.4 (blue to light blue).  There is no systematic variation
in the \sii/\ha\ ratio; for example, the \sii/\ha\ ratio is not
correlated with the surface brightness.

The small \sii/\ha\ ratios within DEM\,L\,192 can also be explained 
by the radiation field provided by the stars in the superbubble.  
Two OB associations are identified within the boundary of DEM\,L\,192: 
LH54 in the central cavity and LH51 near the western rim (Lucke \& 
Hodge 1970; Chu \& Mac Low 1990).  Our WFPC2 field of view is close 
to the OB association LH54, which contains an O4~III(f*) star and a 
WC5 star (\markcite{OS98}Oey \& Smedley 1998).  The O4~III star, 
having an effective temperature of 48,180 K (\markcite{SdK97}Schaerer 
\& de Koter 1997), is among the hottest stars and can easily ionize 
sulfur to S$^{+2}$, thus weakening the \sii\ emission and reducing 
the \sii/\ha\ ratios.  Furthermore, no \hi\ shell is detected 
exterior to the ionized shell of DEM\,L\,192, indicating that the shell 
is optically thin \markcite{kim98b}(Kim et al.\ 1998b; 
\markcite{kim00}Kim et al.\ 2000); therefore, no enhancement to the S$^+$ 
density or the \sii/\ha\ ratios is expected.

\subsection{DEM\,L\,106 (N30C)}

DEM\,L\,106 is an X-ray-dim superbubble; i.e., it has not been detected in 
long-exposure X-ray observations, and the 3$\sigma$ upper limit of its
X-ray luminosity does not exceed that expected by Weaver et al.'s
(1977) bubble models (\markcite{Ch95}Chu et al.\ 1995).  
The expansion velocity of DEM\,L\,106 is $\sim$10 km~s$^{-1}$ 
\markcite{SCG99}(Shaw, Chu, \& Gruendl 2000).
This superbubble contains the OB association LH38 (Lucke \& Hodge 1970);
its stellar content has been studied in detail by \markcite{Oey96a}Oey 
(1996a).

Our WFPC2 \ha\ image covers the shell's center and southern rim.
The morphology of the shell is amorphous with embedded dusty features.
The lack of sharp filaments is consistent with the small expansion
velocity.  At the northern part of the field, in the WF2 CCD, are the 
B[e] star S22 and two compact \hii\ regions.  The \hii\ regions are 
surrounded by a bow-shock-like structure pointing toward S22, suggesting
a possible dynamic interaction.  This interaction will be reported by
Shaw et al.\ (2000).

\subsection{Optical Morphology and X-ray Surface Brightness}

The main purpose for obtaining {\rm HST} WFPC2 images of the 
superbubbles DEM\,L\,152, DEM\,L\,192, and DEM\,L\,106 is to search 
for small, sharp, filamentary, \sii-enhanced features that are 
directly associated with SNR shocks, such as those observed in the 
SNR N63A (Chu et al.\ 1999).  However, these images do not detect
any sharp features with \sii/\ha\ ratios reaching the canonical
value for SNRs, $\ge$0.7.  The absence of detectable SNR-shocked
material indicates that SNRs are not interacting with any
high-density material, such as dense cloudlets or the cold dense 
superbubble shell.  This result supports the previous suggestion 
that the interior SNRs have interacted with only the hot, 
lower-density gas in the superbubbles.

The WFPC2 images show different overall morphologies in the three 
superbubbles, with DEM\,L\,152 being the most filamentary and DEM\,L\,106
the least.  The filamentary morphology of a superbubble is produced
by the post-shock compression as the superbubble expands into the 
ambient interstellar medium.  It is thus expected that superbubbles
with higher expansion velocities should show more filamentary
morphologies.

It has been noted that X-ray-bright superbubbles have higher expansion
velocities than expected in models taking into account realistic
stellar wind energy input \markcite{Oey96b}(Oey 1996b).  Our WFPC2 
images illustrate that the superbubble shell morphologies are caused 
by the interstellar shocks associated with the superbubbles' outward 
expansion, instead of being caused by direct interaction with SNRs.  
There is no morphological or kinematic evidence that interior SNRs in the
X-ray-bright superbubbles DEM\,L\,152 and DEM\,L\,192 shock the cold 
shells directly (this paper; Magnier et al.\ 1996; Kim et al.\ 1998a).
The expanding shells of X-ray-bright superbubbles must have been 
accelerated indirectly.

We suggest that the SNRs in superbubble interiors only shock-heat the 
interior gas, raising the pressure of the superbubble interior.  
The superbubble expansion, driven by an increased pressure, accelerates.
The resultant higher expansion velocities lead to stronger outer 
interstellar shocks, which produce filamentary shell morphologies.   
In this picture, filamentary shell morphologies should be associated
with high X-ray surface brightness of superbubbles.

It is of interest to see whether or not this correlation between 
nebular morphology and X-ray emission is valid in general for 
a large sample of objects.  The {\it HST} archive of WFPC2 \ha\ 
images and the {\it ROSAT} archive of X-ray observations of the 
LMC fields provide excellent data sets for us to examine the 
relationship between nebular morphology and X-ray emission.
We have found in the {\it HST} WFPC2 archive (up to 1999 November)
\ha\ images for $\sim$ 50 fields in the LMC.  Among these \ha\
images, filamentary morphologies comparable to that of DEM\,L\,152 (N44)
are found only in the superbubble N103 around the double cluster
NGC\,1850 (\markcite{Gi94}Gilmozzi et al.\ 1994).  Figure 3a shows 
a WFPC2 \ha\ image of the cluster NGC\,1850 and a portion of the 
superbubble N103.

We have further examined archival {\it ROSAT} Position Sensitive
Proportional Counter observations of N103 (Chu et al. 1999).  As shown 
in Figures 3b and c, N103 is projected near three other X-ray sources:
(1) the bright SNR 0509-68.7 (N103B) outside the northeast rim of the 
superbubble; (2) a point X-ray source at 
5$^{\rm h}$07$^{\rm m}$36$^{\rm s}$, $-68^\circ47'52''$ (J2000),
coincident with the cluster HS122 \markcite{HS66}(Hodge \& Sexton 
1966) projected at the southwest rim of N103; and (3) a large faint
ring of diffuse X-ray emission originating from a probable foreground
SNR in the halo of the LMC (Chu et al.\ 1999).  On the same line of sight,
a fourth component of X-ray emission is present within the superbubble
N103.  The spatial correspondence between this X-ray emission and the 
\ha\ emission suggests a physical association.  Therefore, N103 provides 
another supporting case for the correlation between filamentary 
superbubble morphology and diffuse X-ray emission.  

The existence of a SNR in the interior of N103 has been suggested by 
\markcite{Ac97}Ambrocio-Cruz et al.\ (1997) based on the kinematics of
the superbubble; they find an expansion velocity of 57 km~s$^{-1}$.
The large expansion velocity is probably due to the acceleration of
material by the pressure increase induced by the SNR shock and must 
be responsible for the  filamentary shell morphology.   
These properties are very similar to those of the X-ray-bright 
superbubble DEM\,L\,152 (N44).


\section{Summary and Conclusions}

We have presented $HST$ WFPC2 images of three LMC superbubbles:
DEM\,L\,152, DEM\,L\,192, and DEM\,L\,106.  From these images and all archival
$HST$ WFPC2 images of ionized gas in the LMC, we find that the 
filamentary morphologies of superbubbles are usually associated 
with large expansion velocities and interior X-ray emission.
This correlation may be understood in terms of interior SNRs
near the superbubble shell walls.  SNRs interacting with the 
inner walls of a superbubble may shock-heat and pressurize
the superbubble interior to produce bright X-ray emission and
to accelerate the shell expansion.   Filamentary shell
morphologies are produced by strong interstellar shocks
associated with fast expansion.

We also find that the \sii/\ha\ ratios of superbubbles are 
dependent on the stellar UV radiation field as well as the 
interstellar shocks resultant from the superbubble expansion.
Comparisons between the superbubbles DEM\,L\,152 and DEM\,L\,192 show 
that high \sii/\ha\ ratios are produced only if the ionization 
front is trapped in the superbubble shell.  This conclusion 
differs from \markcite{Oey96b}Oey's (1996b) suggestion that 
high \sii/\ha\ ratios are correlated with large expansion 
velocities and high X-ray surface brightness of superbubbles.
The stellar UV radiation field must be taken into account in 
the interpretation of \sii/\ha\ ratios of superbubbles.

\acknowledgements 
This work was supported by the NASA grant GO-06698.01-95A from
STScI.


\newpage

\centerline{\bf FIGURE CAPTIONS}

\figcaption{$HST$ WFPC2 \ha\ images of (a) DEM\,L\,152, (b) DEM\,L\,192, and (c)
DEM\,L\,106.  The insets are wide-field \ha\ images taken with the CTIO Curtis 
Schmidt telescope; the field of view is 10$'\times10'$.  The WFPC2 
field of view is outlined in the inset.}
\figcaption{[S\,II]/H$\alpha$ ratio maps of N44 and N51D determined from $HST$ 
WFPC2 images.}
\figcaption{(a) $HST$ WFPC2 image of the superbubble N103 around the double
cluster NGC\,1850.  The inset is a wide-field \ha\ image taken with the CTIO 
Curtis Schmidt telescope; the field of view is 10$'\times10'$.  The WFPC2 
field-of-view is outlined in the inset.  (b) $ROSAT$ PSPC X-ray image of N103 
in the 0.5--2.0 keV band, produced from the observations RP300129N00 and 
RP500037N00.  The image has been smoothed by an adaptive-filter with a 
kernal of 50 counts, which is equivalent to a smoothing area of a radius 
of $\sim1'$.
(c) $ROSAT$ PSPC X-ray contours over a CTIO Curtis Schmidt \ha\ image of 
N103.  The contour levels are 0.001, 0.0015, 0.002, 0.0025, 0.005, 0.0075, 
and 0.01 counts s$^{-1}$ arcmin$^{-2}$. }

\newpage

\begin{table}[th]
\caption{Journal of $HST$ WFPC2 Observations}
\begin{tabular}{lccccll}
\hline\hline
Object Name	& R.A.\tablenotemark{a} & Decl.\tablenotemark{a}
&$\lambda_{c} / \Delta\lambda$  &Exposure&Notes&Date of Obs\\
        &   (J2000)           & (J2000)
&(\AA )                        &(s)     &      &\\
\hline
DEM\,L\,152 (N44)	& $5^h$$22^m$$38^s$.5 & -67$^{\circ}$57$\arcmin$47$\arcsec$ &
6563.7/21.4 & 2$\times$500 & \ha & 1999/1/17 \\				   
&&&6732.1/47.2 & 2$\times$600 & \sii & 1999/1/17 \\ 
DEM\,L\,192 (N51D)    & $5^h$$26^m$$15^s$.2 & -67$^{\circ}$29$\arcmin$58$\arcsec$ &
6563.7/21.4 & 2$\times$500 & \ha & 1998/11/16 \\
&&&6732.1/47.2 & 2$\times$600 & \sii & 1998/11/16 \\
DEM\,L\,106 (N30C)& $5^h$$13^m$$41^s$.2 & -67$^{\circ}$27$\arcmin$35$\arcsec$ &
6563.7/21.4 & 3$\times$800 & \ha & 1998/11/14\\
\hline
\tablenotetext{a}{The right ascension and declination of each object 
refers to the center of Planetary Camera (PC1 chip)}
\end{tabular}
\end{table}

\end{document}